\documentclass[preprint,10pt,a4paper,twocolumn,tightenlines]{revtex4}%
\usepackage{amsfonts}
\usepackage{amsmath}
\usepackage{amssymb}
\usepackage{graphicx}%
\begin{document}
\title{Nonadiabatic coherent evolution of two-level systems under spontaneous decay }
\author{F. O. Prado$^{1}$, E. I. Duzzioni$^{1,2}$, M. H. Y. Moussa$^{3}$, N. G. de
Almeida$^{4}$, and C. J. Villas-B\^{o}as$^{1}$}
\affiliation{$^{1}$Departamento de F\'{\i}sica, Universidade Federal de S\~{a}o Carlos,
P.O. Box 676, S\~{a}o Carlos\textit{, }13565-905, S\~{a}o Paulo\textit{, }Brazil}
\affiliation{$^{2}$Centro de Ci\^{e}ncias Naturais e Humanas, Universidade Federal do ABC,
Rua Santa Ad\'{e}lia, 166, Santo Andr\'{e}, S\~{a}o Paulo, 09210-170,\textit{
}Brazil}
\affiliation{$^{3}$Instituto de F\'{\i}sica de S\~{a}o Carlos, Universidade de S\~{a}o
Paulo, Caixa Postal 369, 13560-970, S\~{a}o Carlos, S\~{a}o Paulo, Brazil}
\affiliation{$^{4}$N\'{u}cleo de Pesquisas em F\'{\i}sica, Universidade Cat\'{o}lica de
Goi\'{a}s, 74.605-220, Goi\^{a}nia (GO), Brazil }

\begin{abstract}
In this paper we extend current perspectives in engineering reservoirs by
producing a time-dependent master equation leading to a nonstationary
superposition equilibrium state that can be nonadiabatically controlled by the
system-reservoir parameters. Working with an ion trapped inside a nonindeal
cavity we first engineer effective Hamiltonians that couple the electronic
states of the ion with the cavity mode. Subsequently, two classes of
decoherence-free evolution of the superposition of the ground and decaying
excited levels are achieved: those with time-dependent azimuthal or polar
angle. As an application, we generalise the purpose of an earlier study [Phys.
Rev. Lett. \textbf{96}, 150403 (2006)], showing how to observe the geometric
phases acquired by the protected nonstationary states even under a
nonadiabatic evolution.

\end{abstract}

\maketitle

In the last decade, research activity on the subject of open quantum systems
has been mainly devoted to the search for mechanisms to bypass decoherence.
Beyond the quest for conditions that weaken the system-reservoir coupling
\cite{WC}, coherence control schemes have been introduced among the protocols
for quantum error-correcting codes \cite{QECC}, the existence of
decoherence-free subspaces (DFS) in collective systems \cite{DFS}, and
dynamical decoupling (DD) methods \cite{DD}. More recently, a technique has
been presented to reach the same goal as the DD schemes, without interfering
directly in the system within the reservoir timescale \cite{Lucas}. We finally
mention the engineering reservoir program \cite{Poyatos}, where a quantum
system whose state is to be protected is compelled to engage in additional
interactions besides that with the reservoir. Such interactions are carefully
engineered to modify the Liouvillian in a specific way that drives the system
to an equilibrium with the reservoir. The engineering reservoir has been
developed for both trapped ions \cite{Ions, Matos} and atomic two-level ($TL$)
systems \cite{Atoms}. We stress that the engineering reservoir is deeply
connected to the engineering Hamiltonian program which has been pursued for
quantum information purposes \cite{Fabiano}. More recently, under the
assumption of a squeezed engineered reservoir, a way to observe the adiabatic
geometric phase acquired by a protected state evolving coherently through the
adiabatic manipulation of the squeeze parameters of the engineered reservoir
has been proposed \cite{Carollo2006a, Carollo2006c, Yin 2007}.

In this paper we show how to protect a nonstationary superposition state,
broadening the range of the proposed scheme for engineering reservoirs
\cite{Poyatos}. We present a general recipe to build nonadiabatic coherent
evolutions driven by engineered reservoirs. Given this general recipe, the
task to achieve a particular nonadiabatic evolution --- through a particular
engineered reservoir --- relies entirely on the engineering Hamiltonian
program. Differently from the standard application of the engineering
reservoir technique \cite{Poyatos, Ions, Matos}, in our model we achieve the
nonadiabatic decoherence-free evolution of superposition states, and show how
to implement it in a particular system. Working with an ion trapped inside a
nonideal cavity, we initially engineer an effective Hamiltonian coupling the
electronic levels of the ion with the cavity mode. Assuming a strong decay
rate of the cavity field, this effective interaction is employed to build an
artificial reservoir, leading to two classes of asymptotic nonstationary
superpositions of the ground $\left\vert g\right\rangle $ and decaying excited
$\left\vert e\right\rangle $ ionic levels: those with time-dependent azimuthal
and with time-dependent polar angles. We stress that by combining both
evolutions we can perform any trajectory on the Bloch sphere. As an
application of this engineering reservoir technique, we generalize the
protocol in Ref. \cite{Carollo2006a, Carollo2006b}, demonstrating how to
observe geometric phases acquired by protected nonstationary states even under
a nonadiabatic evolution. As a particular case of our model, we also show how
to build a quantum memory to protect stationary superpositions of the internal
degrees of freedom of the ion.

The main goal of the standard engineering reservoir scheme \cite{Poyatos} is
to obtain, in the interaction picture, a master equation in the form
\begin{equation}
\overset{\cdot}{\rho}=\frac{\Gamma}{2}\left(  2\mathcal{O}\rho\mathcal{O}%
^{\dagger}-\mathcal{O}^{\dagger}\mathcal{O}\rho-\rho\mathcal{O}^{\dagger
}\mathcal{O}\right)  , \label{eqmestra1}%
\end{equation}
where $\Gamma$ is the effective decay rate of the engineered reservoir which
is coupled to the quantum system in a specific way characterized by the
time-independent system operator $\mathcal{O}$. The only pure steady state of
this system is the eigenstate $\left\vert \psi\right\rangle $ of the operator
$\mathcal{O}$ with null eigenvalue, ensuring that there is no further
eigenstate $\left\vert \phi\right\rangle $ of $\mathcal{O}$ such that $\left[
\mathcal{O},\mathcal{O}^{\dagger}\right]  \left\vert \phi\right\rangle =0$
\cite{Matos}. Even considering time-dependent Lindblad operators in master
equation (\ref{eqmestra1}), the scheme used to protect a given state remains
valid, assuming an adiabatic evolution of the reservoir parameters, i. e., the
rate of change of operator $\mathcal{O}$, characterized by $\varpi$, is much
smaller than $\Gamma$. Consequently, we obtain a nonstationary protected state
$\left\vert \psi\left(  t\right)  \right\rangle $, which is the instantaneous
eigenstate of $\mathcal{O}$ with null eingenvalue and follows the adiabatic
changes of the reservoir parameters \cite{Carollo2006b}. Of course, the
fidelity of the protected state in this adiabatic evolution depends on the
ratio $\varpi/\Gamma\ll1$. Next, to remove the adiabatic constraint in the
decoherence-free evolution described above, we consider the engineered
time-dependent master equation in the interaction picture ($\hbar=1$)
\begin{align}
\overset{\cdot}{\rho}  &  =-i\left[  H\left(  t\right)  ,\rho\right]
+\frac{\Gamma}{2}\left[  2\mathcal{O}\left(  t\right)  \rho\mathcal{O}%
^{\dagger}\left(  t\right)  -\mathcal{O}^{\dagger}\left(  t\right)
\mathcal{O}\left(  t\right)  \rho\right. \nonumber\\
&  \left.  -\rho\mathcal{O}^{\dagger}\left(  t\right)  \mathcal{O}\left(
t\right)  \right]  , \label{eqmestra2}%
\end{align}
where the Hermitian Hamiltonian $H\left(  t\right)  $ must be chosen in
accordance with the time dependence of the operator $\mathcal{O}\left(
t\right)  =R\left(  t\right)  \mathcal{O}_{R}R^{\dag}\left(  t\right)  $, with
$R\left(  t\right)  =T\exp\left(  -i\int_{0}^{t}H\left(  t^{\prime}\right)
dt^{\prime}\right)  $, $T$ being the time-ordering operator. Note that through
the unitary transformation $R\left(  t\right)  $, we recover the
time-independent form of the master equation given in (\ref{eqmestra1}), in a
representation where $\mathcal{O}$ is replaced by $\mathcal{O}_{R}$.
Interestingly, the protected stationary eigenstate $\left\vert \psi
_{R}\right\rangle $ ($\mathcal{O}_{R}\left\vert \psi_{R}\right\rangle =0$),
turns out to be a nonstationary state in the original interaction picture,
$\left\vert \psi(t)\right\rangle =R\left(  t\right)  \left\vert \psi
_{R}\right\rangle $, whose evolution can be manipulated by means of
appropriate engineered Hamiltonian $H(t)$ and reservoir.

Now we show how to implement the ideas discussed above, using a $TL$ trapped
ion characterized by the transition frequency $\omega_{0}$ between the ground
$\left\vert g\right\rangle $ and excited $\left\vert e\right\rangle $ states
and trap frequency $\nu$. The transition $\left\vert g\right\rangle $
$\leftrightarrow$ $\left\vert e\right\rangle $ is driven by (one or two)
classical fields of frequencies $\omega_{\ell}$, with coupling strengths
$\Omega_{\ell}$ (with $\ell=1,2$), and the ion is made to interact --- under
the Jaynes-Cummings Hamiltonian and Rabi frequency $g$ --- with a cavity mode
of frequency $\omega_{a}$. Within the rotating-wave approximation ($RWA$), the
Hamiltonian modelling the system is given by
\begin{gather}
H=\omega_{a}a^{\dagger}a+\omega_{0}\sigma_{z}/2+\nu b^{\dagger}b+\left\{
g\cos\left(  \overrightarrow{k}.\overrightarrow{x}\right)  a\sigma
_{eg}\right.  \mathrm{{,}}\nonumber\\
\left.  +\left[  \Omega_{1}\operatorname*{e}\nolimits^{i\left(
\overrightarrow{k_{1}}.\overrightarrow{x}+\phi_{1}-\omega_{1}t\right)
}+\Omega_{2}\operatorname*{e}\nolimits^{i\left(  \overrightarrow{k_{2}%
}.\overrightarrow{x}+\phi_{2}-\omega_{2}t\right)  }\right]  \sigma
_{eg}+\mathrm{H{.c.}}\right\}  \label{1}%
\end{gather}
where $a^{\dagger}$ ($a$) and $b^{\dagger\text{ }}$($b$) are the creation
(annihilation) operators of the quantized harmonic field and the vibrational
mode whose position operator is $\overrightarrow{x}=(b^{\dagger}%
+b)/\sqrt{2m\nu}\widehat{x}$, $m$ being the ionic mass. The wave vectors
$\overrightarrow{k}$, $\overrightarrow{k}_{1}$, and $\overrightarrow{k}_{2}$
stand for the cavity mode and the two amplification fields (with dephasings
$\phi_{1}$ and $\phi_{2}$), respectively, while $\sigma_{kl}\equiv\left\vert
k\right\rangle \left\langle l\right\vert $ ($k$ and $l$ being the states $g$
and $e$). The vibrational mode is decoupled from the remaining degrees of
freedom of our model by assuming the wave vectors $\overrightarrow{k}$,
$\overrightarrow{k}_{1}$, and $\overrightarrow{k}_{2}$ to be perpendicular to
$\overrightarrow{x}$. Under this assumption and going to the interaction
picture through the transformation $U=\exp\left[  -i\left(  \omega
_{a}a^{\dagger}a+\omega_{0}\sigma_{z}/2\right)  t\right]  $, we end up with
the transformed Hamiltonian%
\begin{equation}
H_{1}=\left[  g\operatorname*{e}\nolimits^{-i\Delta_{a}t}a+\Omega
_{1}\operatorname*{e}\nolimits^{i\left(  \phi_{1}-\Delta_{1}t\right)  }%
+\Omega_{2}\operatorname*{e}\nolimits^{i\left(  \phi_{2}-\Delta_{2}t\right)
}\right]  \sigma_{eg}+\mathrm{H{.c.}}{,} \label{H1}%
\end{equation}
where we have defined the detunings $\Delta_{a}=\omega_{a}-\omega_{0}$ and
$\Delta_{\ell}=\omega_{\ell}-\omega_{0}$.

\textit{Nonadiabatic decoherence-free evolution }. To accomplish a
decoherence-free evolution of a superposition of the atomic levels, we must
first engineer the appropriate interaction between these levels and the cavity
mode. To this end we have to adjust the first classical field to resonance
with the atomic transition, i.e., $\Delta_{1}=0$. In what follows, we perform
two consecutive unitary transformations, first to a framework rotating with
frequency $\Omega_{1}$, $U_{1}=\exp\left[  -i\Omega_{1}\left(
\operatorname*{e}\nolimits^{i\phi_{1}}\sigma_{eg}+\mathrm{H{.c.}}\right)
t\right]  $, which is straightforwardly done with the help of the basis states
$\left\{  \left\vert \pm\right\rangle =\left(  \left\vert e\right\rangle
\pm\operatorname*{e}\nolimits^{-i\phi_{1}}\left\vert g\right\rangle \right)
/\sqrt{2}\right\}  $, constituting the eigenstates of the operator $\Omega
_{1}\left(  \operatorname*{e}\nolimits^{i\phi_{1}}\sigma_{eg}+\mathrm{H{.c.}%
}\right)  $ defining $U_{1}$. The adjustment $\Delta_{2}=-2\Omega_{1}$ enables
us to proceed to the second transformation $U_{2}=\exp\left[  -i\Omega
_{2}\left(  \operatorname*{e}\nolimits^{i\varphi}\sigma_{+-}+\mathrm{H{.c.}%
}\right)  t/2\right]  $, performed with the help of another set of basis
states $\left\{
\genfrac{\vert}{\rangle}{0pt}{}{\uparrow}{\downarrow}%
=\left(  \left\vert +\right\rangle \pm\operatorname*{e}\nolimits^{-i\varphi
}\left\vert -\right\rangle \right)  /\sqrt{2}\right\}  $, composed by
eigenstates of the operator $\Omega_{2}\left(  \operatorname*{e}%
\nolimits^{i\varphi}\sigma_{+-}+\mathrm{H{.c.}}\right)  $. Working in the
regime where $\Omega_{1}\sim\Delta_{2}\gg$ $\Omega_{2}\sim\Delta_{a}\gg g$ and
adjusting $\Delta_{a}=-\Omega_{2}$, we end up, after a $RWA$, with the
effective Hamiltonian%
\begin{equation}
H_{2}=\left(  g/2\right)  \left(  \operatorname*{e}\nolimits^{i\phi_{1}%
}a^{\dagger}\sigma_{\uparrow\downarrow}+\operatorname*{e}\nolimits^{-i\phi
_{1}}a\sigma_{\downarrow\uparrow}\right)  \mathrm{{.}} \label{H2}%
\end{equation}

Now, with the engineered interaction (\ref{H2}) and the dissipative mechanisms
of both the harmonic mode and the $TL$ system, the evolution of the
transformed density operator $\rho$ of the whole system is given by%

\[
\overset{\cdot}{\rho}=-i\left[  H_{2}\mathrm{{,}}\rho\right]  +\left(
\Gamma/2\right)  \left(  2a\rho a^{\dagger}-a^{\dagger}a\rho-\rho a^{\dagger
}a\right)  +\mathcal{L}_{TL}\rho{,}%
\]
where $\mathcal{L}_{TL}\rho$ stands for the Liouvillian dynamics of the $TL$
system under the transformations $U_{1,2}$ which do not modify the usual
Liouvillian form for the harmonic field decay\textbf{.} Towards the engineered
reservoir, we assume that the decay constant of the cavity field is
significantly larger than both the effective coupling $g/2$ and the decay
constant $\gamma$ of the $TL$ system in $\mathcal{L}_{TL}\rho$. In our
\textquotedblleft cavity QED + trapped ion\textquotedblright\ system, the
regime $\Gamma\gg g$, $\gamma$ is easily achieved through a cavity with low
quality factor $Q=\omega_{a}/\Gamma$. Together with the good approximation of
a reservoir at absolute zero, this regime enables us to consider only the
matrix elements $\rho_{mn}=\left\langle m\right\vert \rho\left\vert
n\right\rangle $ inside the subspace $\left\{  \left\vert 0\right\rangle
\mathrm{{,}}\left\vert 1\right\rangle \right\}  $ of photon numbers. Moreover,
following the reasoning in Ref. \cite{Matos}, the strong decay rate $\Gamma$
enables the adiabatic elimination of the elements $\rho_{01}$ and $\rho_{11}$,
prompting the evolution of the $TL$ system%
\begin{equation}
\overset{\cdot}{\rho}_{TL}=\Gamma_{eng}\left(  2\sigma_{\uparrow\downarrow
}\rho_{TL}\sigma_{\downarrow\uparrow}-\sigma_{\downarrow\downarrow}\rho
_{TL}-\rho_{TL}\sigma_{\downarrow\downarrow}\right)  +\mathcal{L}_{TL}%
\rho_{TL}\mathrm{{,}} \label{Rho}%
\end{equation}
where $\Gamma_{eng}=g^{2}/\Gamma$ stands for the coupling strength of the
engineered reservoir. The inevitable and undesired action of the multimode
vacuum $\mathcal{L}_{TL}\rho$ thus works against the protected state
$\left\vert \uparrow\right\rangle \left\langle \uparrow\right\vert $ of the
$TL$ system which follows asymptotically from Eq. (\ref{Rho}) with $\gamma=0$.
In fact, taking into account the multimode vacuum, the equations of motion of
the matrix elements $\left(  \rho_{TL}\right)  _{rs}=\left\langle r\right\vert
\rho\left\vert s\right\rangle $ (with $r$ and $s$ standing for $\uparrow$ and
$\downarrow$), following from the rotating-wave approximation, are given by%
\begin{align*}
\left(  \overset{\cdot}{\rho}_{TL}\right)  _{\uparrow\uparrow}  &  =\left(
\Gamma_{eng}+3\gamma/8\right)  -\left(  \Gamma_{eng}+6\gamma/4\right)  \left(
\rho_{TL}\right)  _{\uparrow\uparrow}\\
&  =-\left(  \overset{\cdot}{\rho}_{TL}\right)  _{\downarrow\downarrow
}\text{,}\\
\left(  \overset{\cdot}{\rho}_{TL}\right)  _{\uparrow\downarrow}  &  =-\left(
\Gamma_{eng}/2+5\gamma/4\right)  \left(  \rho_{TL}\right)  _{\uparrow
\downarrow}+\left(  \gamma/8\right)  \left(  \rho_{TL}\right)  _{\downarrow
\uparrow}\\
&  =\left(  \overset{\cdot}{\rho}_{TL}\right)  _{\downarrow\uparrow}^{\ast
}\text{,}%
\end{align*}
whose asymptotic solution leads to the protected state
\begin{equation}
\rho_{TL}(t\rightarrow\infty)=\left(  1-\varepsilon\right)  \left\vert
\uparrow\right\rangle \left\langle \uparrow\right\vert +\varepsilon\left\vert
\downarrow\right\rangle \left\langle \downarrow\right\vert \text{,}
\label{RhoP}%
\end{equation}
where $\varepsilon=\left[  2+\left(  8/3\right)  \left(  \Gamma_{eng}%
/\gamma\right)  \right]  ^{-1}$. From Eq. (\ref{RhoP}), it is straightforward
to compute the fidelity of the protected state $\left\vert \uparrow
\right\rangle $, given by $\mathcal{F}=\operatorname*{Tr}\left[  \left\vert
\uparrow\right\rangle \left\langle \uparrow\right\vert \rho_{TL}%
(t\rightarrow\infty)\right]  =1-\varepsilon$, which approaches unity for a
successfully engineered coupling strength $\Gamma_{eng}\gg\gamma$. It is worth
noting that even the modest ratio $\Gamma_{eng}/\gamma=10$ results in a
fidelity around $97\%$. Considering that the coupling between the ground and
excited states is induced by a Raman transition \cite{Ions}, where
$g\approx10^{5}$ s$^{-1}$ and $\gamma\approx10^{2}$ s$^{-1}$
\cite{Matos,Serra}, we obtain for a cavity decay constant $\Gamma\approx
10^{6}$ s$^{-1}$ the strength $\Gamma_{eng}/\gamma\approx10^{2}$. Therefore,
with the excellent approximation $\varepsilon\ll1$ and reversing the unitary
transformations $U_{2}$ and $U_{1}$, we note that the state $\left\vert
\uparrow\right\rangle $, written in the interaction picture as%
\begin{equation}
\left\vert \psi(t)\right\rangle =\cos\left(  \varphi/2-\Omega_{1}t\right)
\left\vert e\right\rangle -i\operatorname*{e}\nolimits^{-i\phi_{1}}%
\sin(\varphi/2-\Omega_{1}t)\left\vert g\right\rangle \text{,} \label{SP1}%
\end{equation}
allows for a nonadiabatic\ coherent evolution of a $TL$ system under
spontaneous decay, which can be manipulated through the laser parameters
$\Omega_{1}$, $\varphi=\phi_{1}-\phi_{2}$. Such an evolution corresponds to
flips in the atomic states, representing trajectories on different meridian
planes on the Bloch sphere, governed by the unitary evolution $\exp\left\{
-i\left[  \left(  \Omega_{1}t-\varphi/2\right)  \left(  \sigma_{eg}%
\operatorname*{e}\nolimits^{i\phi_{1}}+\mathrm{{H.c.}}\right)  \right]
\right\}  $. At this point we note that the time-independent operator
$\sigma_{\uparrow\downarrow}$ corresponds to $\mathcal{O}_{R}$ mentioned
above, while the transformation $U_{2}U_{1}$ corresponds to $R$.

\textit{Quantum memory and adiabatic decoherence-free evolution}. To build a
quantum memory, a device that protects a stationary superposition, we turn off
the second amplification field ($\Omega_{2}=0$) in our starting Hamiltonian
(\ref{H1}). Applying the unitary transformation $\widetilde{U}_{1}=\exp\left[
i\Delta_{1}\sigma_{z}/2t\right]  $, we thus obtain the time-dependent
Hamiltonian $\widetilde{H}_{1}=\Delta_{1}\sigma_{z}/2+\Omega_{1}\left(
\operatorname*{e}\nolimits^{i\phi_{1}}\sigma_{eg}+\operatorname*{e}%
\nolimits^{-i\phi_{1}}\sigma_{ge}\right)  +\left[  g\operatorname*{e}%
\nolimits^{-i\Delta_{a}t}a\sigma_{eg}+\mathrm{H{.c.}}\right]  $. Under the
additional transformation $\widetilde{U}_{2}=\exp\left\{  -i\left[  \Delta
_{1}\sigma_{z}/2+\Omega_{1}\left(  \operatorname*{e}\nolimits^{i\phi_{1}%
}\sigma_{eg}+\operatorname*{e}\nolimits^{-i\phi_{1}}\sigma_{ge}\right)
\right]  t\right\}  $, accomplished with the help of the basis states
$\left\{  \left\vert \widetilde{\pm}\right\rangle =\left(  \sqrt{2\pm\chi
}\left\vert e\right\rangle \pm\operatorname*{e}\nolimits^{-i\phi_{1}}%
\sqrt{2\mp\chi}\left\vert g\right\rangle \right)  /2\right\}  $, with
$\chi=\Delta_{1}/\lambda$, $\lambda=\sqrt{\Omega_{1}^{2}+\Delta_{1}^{2}/4}$,
and the detuning $\Delta_{a}=-2\lambda$, we finally obtain the effective
interaction%
\begin{equation}
\widetilde{H}_{2}=(\widetilde{g}/2)\left(  \operatorname*{e}\nolimits^{i\phi
_{1}}a^{\dagger}\sigma_{+-}+\operatorname*{e}\nolimits^{-i\phi_{1}}%
a\sigma_{-+}\right)  {,}%
\end{equation}
where $\widetilde{g}=g(1-\chi/2)$.

Following the steps outlined above for the addition of the damping mechanism
to the cavity mode and the $TL$ system, we reach the master equation
$\overset{\cdot}{\widetilde{\rho}}_{TL}=\widetilde{\Gamma}_{eng}\left(
2\sigma_{+-}\widetilde{\rho}_{TL}\sigma_{-+}-\sigma_{--}\widetilde{\rho}%
_{TL}-\widetilde{\rho}_{TL}\sigma_{--}\right)  +\widetilde{\mathcal{L}}%
_{TL}\widetilde{\rho}_{TL}$, with $\widetilde{\Gamma}_{eng}=\widetilde{g}%
^{2}/\Gamma$, leading to the asymptotic solution%
\begin{align*}
\widetilde{\rho}_{TL}(t  &  \rightarrow\infty)=\left(  1-\widetilde
{\varepsilon}\right)  \left\vert +\right\rangle \left\langle +\right\vert
+\widetilde{\varepsilon}\left\vert -\right\rangle \left\langle -\right\vert \\
&  +\widetilde{\varepsilon}\left(  1-\widetilde{\varepsilon}\right)
^{-1}\left(  \left\vert +\right\rangle \left\langle -\right\vert +\left\vert
-\right\rangle \left\langle +\right\vert \right)  \text{,}%
\end{align*}
where $\widetilde{\varepsilon}=\left[  2+\left(  \widetilde{\Gamma}%
_{eng}/\gamma\right)  \right]  ^{-1}\ll1$ for the case where $\widetilde
{\Gamma}_{eng}\gg\gamma$, providing again a fidelity ($\mathcal{F}%
=1-\widetilde{\varepsilon}$) around unity for the protected state $\left\vert
+\right\rangle $. For the approximation ($\widetilde{\varepsilon}\ll1)$, the
state $\left\vert +\right\rangle $ written in the interaction picture is given
by%
\begin{equation}
\left\vert \widetilde{\psi}(t)\right\rangle =\frac{\sqrt{2+\chi}\left\vert
e\right\rangle +\operatorname*{e}\nolimits^{-i\left(  \phi_{1}-\Delta
_{1}t\right)  }\sqrt{2-\chi}\left\vert g\right\rangle }{2}\text{.} \label{SP2}%
\end{equation}
In the resonant case, where $\Delta_{1}=\chi=0$, we obtain the stationary
state $\left\vert +\right\rangle =\left(  \left\vert e\right\rangle
+\operatorname*{e}\nolimits^{-i\phi_{1}}\left\vert g\right\rangle \right)
/\sqrt{2}$, which would have been its complementary state $\left\vert
-\right\rangle $ if we had set the detuning $\Delta_{a}=2\lambda$ or the phase
$\phi_{1}\rightarrow\phi_{1}+\pi$. In that case, it is worth noting that if
the value of the dephasings changes adiabatically between $\phi_{1}$ and
$\phi_{1}+2\pi$, in such a way that the system is always in equilibrium with
the engineered reservoir, the Bloch vector representing the protected state
(\ref{SP2}) performs a complete rotation around the Bloch sphere, as required
in Ref. \cite{Carollo2006a}, to achieve a coherent evolution of a
superposition state driven by an engineered reservoir.

From the above definition, we note that the ratio $\widetilde{\Gamma}%
_{eng}/\gamma=\left[  g(1-\chi/2)\right]  ^{2}/\left(  \gamma\Gamma\right)  $
is a function of the parameter $\chi$ which defines the polar angle of the
state vector (\ref{SP2}) on the Bloch sphere. This ratio reaches a maximum
when $\Omega_{1}/\Delta_{1}\rightarrow0$ with negative $\Delta_{1}$ and,
consequently, $\chi=-2$, corresponding to the ground state $\left\vert
g\right\rangle $. When $\Omega_{1}/\Delta_{1}\rightarrow0$ with positive
$\Delta_{1}$, such that $\chi=2$, the ratio $\widetilde{\Gamma}_{eng}/\gamma$
is null, forbidding us from protecting the excited state $\left\vert
e\right\rangle $. Evidently, the values $\Delta_{1}$,$\chi\neq0$ describes
nonadiabatic evolution on the Bloch sphere. For the intermediate value
$\chi=0$, we obtain from the typical strengths considered above for $g$,
$\gamma$, and $\Gamma$ the value $\widetilde{\Gamma}_{eng}/\gamma=10^{2}$,
representing a fidelity around $99\%$ for the protected `equatorial' state
$\left\vert +\right\rangle $. Whereas the `equatorial' case $\Delta_{1}%
=\chi=0$ can be employed to achieve an adiabatic azimuthal evolution of the
stationary state $\left\vert +\right\rangle =\left(  \left\vert e\right\rangle
+\operatorname*{e}\nolimits^{-i\phi_{1}}\left\vert g\right\rangle \right)
/\sqrt{2}$ on the Bloch sphere, the cases $\Delta_{1}$,$\chi\neq0$ describe
nonadiabatic evolution on the parallel planes on the Bloch sphere.

\textit{Geometric phase induced by reservoir. }As can be seen from Eqs.
(\ref{SP1}) and (\ref{SP2}), in the context of decoherence-free evolution, we
are able to engineer the non-stationary superposition of atomic states,
evolving coherently and acquiring geometric and dynamic phases. In fact,
rewriting the state (\ref{SP1}) as $\left\vert \psi(t)\right\rangle =\left[
\left\vert +\right\rangle +e^{-i\left(  \varphi-2\Omega_{1}t\right)
}\left\vert -\right\rangle \right]  /\sqrt{2}$, we obtain after a cyclic
evolution ($T=\pi/\Omega_{1}$) the dynamic phase $\phi_{D}(T)=-\int_{0}%
^{T}\left\langle \psi(t)\right\vert H_{I}(t)\left\vert \psi(t)\right\rangle
dt=-\pi\Omega_{2}/\left(  2\Omega_{1}\right)  $ and the geometric one
$\phi_{G}^{cic.}(T)=i\int_{0}^{T}\left\langle \psi(t)\right\vert \frac{d}%
{dt}\left\vert \psi(t)\right\rangle dt=-\pi$, respectively, where
$H_{I}(t)=\Omega_{1}\left(  \sigma_{++}-\sigma_{--}\right)  +\Omega_{2}\left[
e^{i\left(  \varphi-2\Omega_{1}t\right)  }\sigma_{+-}+e^{-i\left(
\varphi-2\Omega_{1}t\right)  }\sigma_{-+}\right]  /2$. Therefore, considering
the total evolution time $T$, under the regime of parameters stated above
($\Omega_{2}\ll\Omega_{1}$), we see that the contribution coming from the
dynamic phase is negligible, while the geometric phase is $\pi$.

In order to observe geometric effects we consider an auxiliary atomic level
$a$, which does not couple with the states $\left\vert g\right\rangle $ and
$\left\vert e\right\rangle $ through the action of the fields involved in the
engineering scheme. We observe that within the time scale $T$ of the
experiment, the lifetime of state $\left\vert a\right\rangle $ does not affect
the dynamics described by the master equation (\ref{Rho}).\ Otherwise the
level $a$ can be chosen as a more excited metastable state. To measure the
phases acquired by the state $\left\vert \psi(t)\right\rangle $, we must
employ an interferometric scheme with $\left\vert a\right\rangle $ as the
reference state \cite{Carollo2006a}. For this purpose, using the conservation
of the total probability, $\rho_{aa}^{I}(t)+\rho_{++}^{I}(t)+\rho_{--}%
^{I}(t)=1$, we solve the\ system of coupled differential equations for the
probability amplitudes $\rho_{ij}^{I}(t)$ ($i,j=+,-,a$) following from Eq.
(\ref{Rho}) in the interaction picture, disregarding, within the evolution
time $T$, the small contribution of $\mathcal{L}_{TL}$. Supposing now that the
initial state of the system is $\left\vert \Psi_{I}(0)\right\rangle =\left(
\left\vert \psi(0)\right\rangle +\left\vert a\right\rangle \right)  /\sqrt
{2},$with $\phi_{1}=\phi_{2}=0$, we find at time $t$ that $\rho^{I}%
(t)=\left\vert \Psi_{I}(t)\right\rangle \left\langle \Psi_{I}(t)\right\vert $,
where $\left\vert \Psi_{I}(t)\right\rangle =\left(  \left\vert a\right\rangle
+e^{-i\left(  \Omega_{1}+\Omega_{2}/2\right)  t}\left\vert \psi
(t)\right\rangle \right)  /\sqrt{2}$. As we have set $\phi_{1}=\phi_{2}=0$,
the protected state at $t=0$ turns out to be $\left\vert \psi(0)\right\rangle
=\left\vert e\right\rangle $. For this reason, the superposition state
$\left\vert \Psi_{I}(0)\right\rangle $ may be obtained by applying a laser
pulse between the states $\left\vert a\right\rangle $ and $\left\vert
e\right\rangle $. Finally, the geometric phase may be observed through the
population inversion $P_{ea}(t)=\cos\left[  \left(  2\Omega_{1}+\Omega
_{2}\right)  t\right]  /2\simeq\cos\left[  2\phi_{G}^{cic.}(t)\right]  /2$,
where $t=n\pi/\Omega_{1};n\in%
\mathbb{N}
$.

It is worth noting that, differently from the scheme proposed in Ref.
\cite{Carollo2006a}, where the superposition $\left\vert \Psi_{I}\left(
t\right)  \right\rangle $, used to measure the geometric phase, is affected by
decoherence under a nonadiabatic evolution, here $\left\vert \Psi\left(
t\right)  \right\rangle $ is unaffected by the reservoir even under such an
evolution faster than that determined by the time scale of the engineered reservoir.

Summarizing, we have improved the engineering reservoir by producing a
time-dependent master equation leading to a nonstationary superposition
equilibrium state that can be nonadiabatically controlled by the
system-reservoir parameters. Working with an ion trapped inside a bad cavity
we constructed two classes of decoherence-free evolution of the ground and
excited ionic levels. By combining the two classes of evolution, we can
manipulate trajectories on the Bloch sphere by changing, alternately, the
polar and azimuthal angles. Although in our schemes the protected states
acquire dynamic phases, this fact is unimportant, since they remain in a
decoherence-free subspace, where the dynamic and geometric phases are
unaffected by the reservoir. Finally, we have also generalized the objective
of the Refs. \cite{Carollo2006a, Carollo2006b}, showing how to observe the
geometric phases acquired by the protected nonstationary states even under a
nonadiabatic evolution. We believe that the extention of the present scheme
for the nonadiabatic time-dependent control of a set of qubits, generating
quantum logic operations inside decoherence-free subspaces, may improve
quantum computation.

We wish to express our thanks to the UFABC and for the support from FAPESP,
CAPES, and CNPq, Brazilian agencies.

\end{document}